\documentclass[
twocolumn,
superscriptaddress,
amsmath,amssymb,
prl,
]{revtex4-2}
\usepackage{graphicx}
\usepackage{hyperref}
\usepackage{dcolumn}
\usepackage{bm}
\usepackage{epstopdf}
\usepackage{romannum}
\usepackage{csquotes}
\usepackage[dvipsnames]{xcolor}
\usepackage{hyperref}

\begin{document}
	
	\preprint{PbTaSe2}
	
	\title{Effects of disorder on the quantum transport porperities in topologically nontrivial metal PbTaSe$_{2}$}
	
	\author{Longfei Sun}
	\affiliation{Key Laboratory of Quantum Materials and Devices of Ministry of Education, School of Physics, Southeast University, Nanjing 211189, China}
	
	\author{Yue Sun}
	\email{Corresponding author:sunyue@seu.edu.cn}
	\affiliation{Key Laboratory of Quantum Materials and Devices of Ministry of Education, School of Physics, Southeast University, Nanjing 211189, China}
	
	\author{Qiang Hou}
	\affiliation{Key Laboratory of Quantum Materials and Devices of Ministry of Education, School of Physics, Southeast University, Nanjing 211189, China}

	\author{R. Sankar}
     \affiliation{Institute of Physics, Academia Sinica, Taipei 10617, Taiwan}

    \author{R. Kalaivanan}
    \affiliation{Institute of Physics, Academia Sinica, Taipei 10617, Taiwan}     

	\author{Xiaofeng Xu}
    \email{Corresponding author:xuxiaofeng@zjut.edu.cn}
    \affiliation{Department of Applied Physics, Zhejiang University of Technology, Hangzhou, 310023, China}	

	\author{Zhixiang Shi}
	\email{Corresponding author:zxshi@seu.edu.cn}
	\affiliation{Key Laboratory of Quantum Materials and Devices of Ministry of Education, School of Physics, Southeast University, Nanjing 211189, China}
	
	\author{Tsuyoshi Tamegai}
	\affiliation{Department of Applied Physics, The University of Tokyo, Tokyo 113-8656, Japan}

	\begin{abstract}
		
		\textbf{}
			Weak antilocalization (WAL), an increase in the electrical conductivity at low temperatures associated with the suppression of electron localization due to quantum interference effects, is often observed in topological materials. In this study, we report the observation of WAL in topologically nontrivial metal PbTaSe$_{2}$ at low temperatures. In the pristine sample, we identified the presence of WAL, which is attributed to the topologically protected backscattering. In order to investigate the influnce of disorder to the WAL, we successively introduced controlled amounts of disorder by H$^{+}$-irradiation. As disorder increases, the dip-like magnetoresistance caused by WAL changes to a linear magnetoresistance(MR), and eventually to a quadratic MR as the electronic system becomes highly localized. This research unveils the significance of disorder in shaping the quantum transport characteristics of topological materials.
	\end{abstract}
	
	\maketitle
	\textbf{}
       Weak antilocalization (WAL) is a fascinating quantum interference effect that arises in disordered electronic systems when the interference between time-reversed electron trajectories leads to an enhancement of conductivity \cite{66,15,67}. This phenomenon is in stark contrast to the classical prediction of decreasing conductivity with decreasing temperature due to Anderson localization \cite{68,69}. WAL is particularly relevant in the study of mesoscopic and nanoscale systems, where quantum coherence effects become significant \cite{59,60}. On the other hand, electrons in a disordered material can be scattered by impurities and defects. Quantum mechanically, the electron wave functions can interfere constructively or destructively, affecting the position of the cruising electrons. \cite{70,74,20}. In the case of WAL, the constructive interference of time-reversed paths through localized states leads to an increased probability of electron transmission and thus higher conductivity. The key factors that contribute to the emergence of WAL include the presence of time-reversal symmetry, spin-orbit coupling, and quantum coherence effects. Time-reversal symmetry alters the interference patterns of electron wavefunctions, leading to an enhancement of conductivity in disordered conductors. Spin-orbit coupling introduces a phase shift between electron spin states, altering the interference pattern and affecting the conductivity \cite{71}. Lastly, quantum coherence refers to the maintenance of phase relationships between different quantum states. Phase coherence is essential for constructive interference between multiple scattering paths, leading to an enhanced conductivity.  Additionally, WAL can be influenced by external factors such as magnetic field, which modifies the interference patterns and enable experimental observation \cite{28,29,30}.
       
       WAL has already been observed in graphene and some topological insulators, which also leads us to think about another class of materials that share similar topological properties \cite{61,62,63}. Topological superconductors (TSCs) are materials with a novel quantum state of matter characterized by nontrivial topological features in their Cooper pairing states. TSCs  are distinct from conventional superconductors hosting exotic phenomena such as Majorana zero modes, which are non-Abelian anyons. These unusual properties open up new possibilities for quantum computation and fault-tolerant quantum information processing. The rapid development of topological materials has provided new avenues for intrinsic TSCs. For instance, chemical doping of topological insulators (TIs) has presented a promising route to TSCs, leading to the discovery of a 3D TSC candidate Cu$_{x}$Bi$_{2}$Se$_{3}$ with a transition temperature $T_{\rm{c}}$$\sim$3.8 K \cite{2}. Additionally, PbTaSe$_{2}$ was found to have a topological surface state with a fully gapped bulk superconductivity \cite{13,57}. It is a noncentrosymmetric superconductor, characterized by the alternating stacking of hexagonal TaSe$_{2}$ and Pb layers, with a space group of $P$$\bar{6}$m2 ($No$.187) \cite{12}. The superconducting transition temperature is 3.8 K. Theoretical calculations indicate a nonzero $Z$$_{2}$ topological invariant in the band structure, leading to spin-helical surface states near the Fermi level. These distinct surface state has been observed in angle-resolved photoemission spectroscopy \cite{11}.
    
    \begin{figure*}\center
    	\includegraphics[width=0.8\linewidth, height=0.36\textheight]{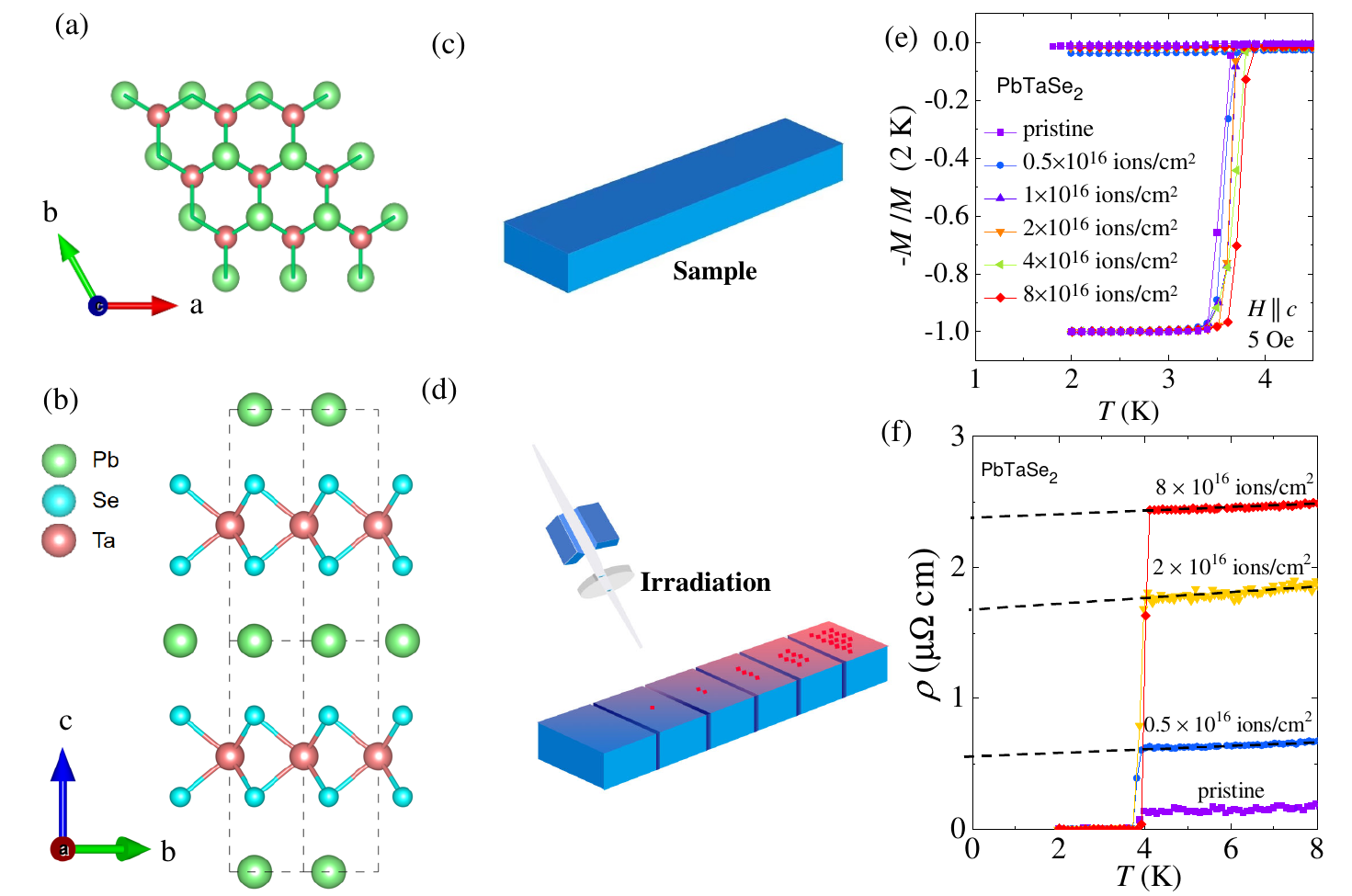}
 	    \caption{The crystal structure of PbTaSe$_{2}$ in (a) the hexagonal plane from the top view, and (b) the noncentrosymmetric structure from the side view. The green, red, and blue circles represent the elements of Pb, Ta, and Se, respectively. (c,d) a piece of single crystal is divided into several pieces and were conducted irradiation experiments separately. Temperature dependence of the (e) normalized magnetization and (f)  resistivity for the pristine and several selected irradiated crystals.}\label{}
    \end{figure*}
    
        In this article, we studied the WAL in topological superconductor candidate PbTaSe$_{2}$ with controllable amounts of disoder created by H$^{+}$-irradiation. With the increase of disorder, the WAL was found to be gradulally suppressed. Furthermore, our research unveils the fundamental mechanisms behind the observed quantum transport behavior in PbTaSe$_{2}$, providing valuable insights into how topological materials harness and manipulate quantum effects. 

        High-quality PbTaSe$_{2}$ single crystals were grown by the vapor transport method (CVT) \cite{23}. The obtained crystals show high quality with sharp superconducting transition width $<$0.5 K from susceptibility measurements, and large residual resistivity ratio [RRR =$\rho$(300 K)/$\rho$(5 K)] $\sim$ 139, as reported in our previous work \cite{24}. Transport measurements were performed by using the four-lead method with a physical property measurement system (PPMS, Quantum Design). Single crystals used for the irradiation experiments were cleaved to thin plates with thickness $\sim$ 30 $\mu$m along the $c$ axis, which is smaller than the projected range of 3-MeV H$^{+}$ for PbTaSe$_{2}$ of $\sim$ 47 $\mu$m. To avoid possible sample-dependent influence, all the measurements were performed on one identical piece of crystal, which was divided into several pieces and irradiated by H$^{+}$ up to doses of 0 (pristine), 0.5$\times$10$^{16}$, 1$\times$10$^{16}$, 2$\times$10$^{16}$, 4$\times$10$^{16}$, and 8$\times$10$^{16}$/cm$^{2}$, respectively (see Figs. 1(c)-(d)). We estimated that 1$\times$10$^{16}$ dose H$^{+}$ irradiation is supposed to cause about 1 point defect per 3200 Pb atoms assuming no overlap. More details about the irradiation experiments can be seen in our previous publications \cite{25,26,27,28}.
	
	    PbTaSe$_{2}$ exhibits a unique crystal structure characterized by the stacking of hexagonal TaSe$_{2}$ layers with intercalated Pb atoms. Figure 1(a) provides an overview of the crystal structure when viewed along the hexagonal TaSe$_{2}$ plane. In this arrangement, the Pb atoms are positioned above the Se atoms, contributing to the crystal's noncentrosymmetric nature, as depicted in the side view presented in Fig. 1(b). Figs. 1(e) and 1(f) show the resistivity and magnetization of PbTaSe$_{2}$ crystals as functions of temperature. Clearly, the superconducting transition temperature ($T{_{c}}$) remains nearly unchanged after irradiation. The unchanged $T$$_{c}$ suggests the absence of nodes in the superconducting gap structure, which has been discussed in detail in our previous report \cite{24}. On the other hand, the enhancement of the residul resistivity after irradiation confirms the introduce of disorder, as shown by the dashed line in Fig. 2(b).
	
	 \begin{figure}
		\centering
		\includegraphics[width=1.0\linewidth]{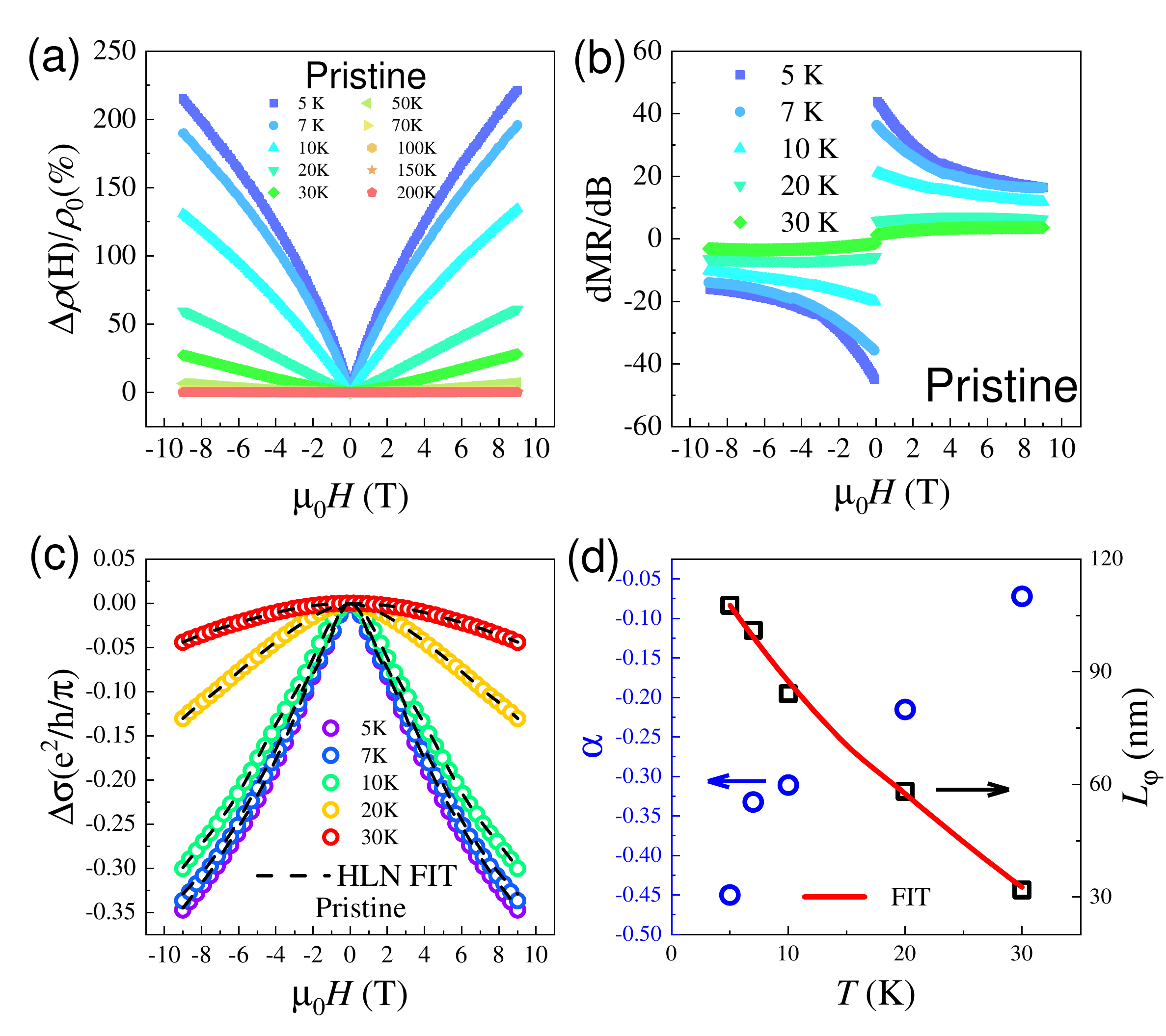}
		\caption{(a) The magnetoresistance measured at different temperatures indicated with applied magnetic field perpendicular to the $ab$ plane.
			(b) First order derivative of MR as a function of magnetic field.(c) The fitting curves using the modified HLN model. (d) Temperature dependence of the fitting parameters $\alpha$ (left axis) and $L$$_\phi$ (right axis).}
		\label{fig:fig2}
	\end{figure}
	
     	To study the irradiation effects on the transport properties of our crystals, we measured the MR at some selected crystals. The MR is defined as MR$\%$= [$\rho$($B$) $-$ $\rho$(0)]/$\rho$(0)$\times$100$\%$, where $\rho$($B$) is the resistivity at the applied magnetic field $B$, and $\rho$(0) is the measured resistivity at $B$ = 0. Evidently, the MR for the pristine single crystal (see Fig. 2(a)) shows a large value, over 200$\%$ at 9 T and 5 K which is consistent with previous reports \cite{31}. In particular, at low temperatures, it exhibits a V-shaped MR instead of a normal U-shape. Besides, a non-saturating MR which can be approximately described by MR($B$) $\propto$ $B$$^\alpha$ ($\alpha$ $\leq$ 1) is observed for the entire temperature and field range. Upon closer examination of the low-field region of the MR, a sharp MR dip is observed (Fig. 2a), indicating the presence of a WAL effect. As the temperature increases, the dip broadens, suggesting a suppression of localizationin, which is attributed to the decrease in phase coherence length \cite{46}, and will be discussed in more detail below.

		\begin{figure*}\center
		\includegraphics[width=1.0\linewidth]{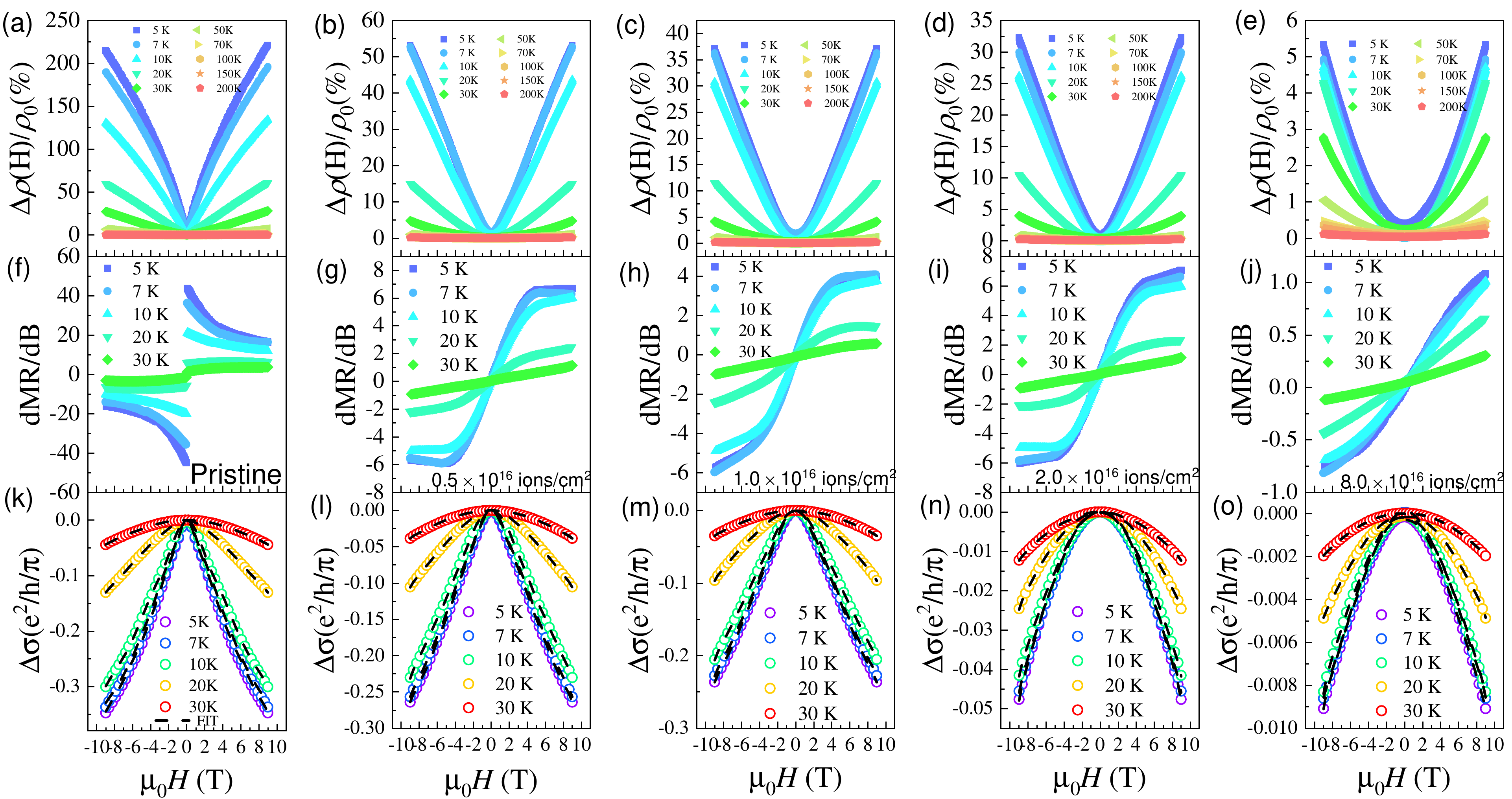}
		\caption{(a)-(d) Morphological changes of magnetoresistance (MR) for samples with different irradiation concentrations. (e)-(h) Variations in the derivative of MR corresponding to different samples. (i)-(l) Variations in MC of different irradiated samples with respect to magnetic field, i.e., variations in the degree of WAL.}\label{}
	    \end{figure*}
	  
       On the other hand, the MR turns to be linearly increased with fields in the high-field range (with $\alpha$$\sim$1). It can be seen more clearly in the Fig. 2(b), where dMR/dB becomes a constant under high field. Many topological materials exhibit a common magnetotransport characteristic where a WAL effect is observed at low temperatures, accompanied by a non-saturating linear-like MR at higher temperatures \cite{32,64,65}. In past studies, the linear-like MR can be attributed to the Dirac bands \cite{49} which is explained by Abrikosov’s quantum linear MR mechanism \cite{72}. The quantum oscillation has been observed by the thermoelectric measurements of seebeck and torgue signals at the fields where the linear MR is observed. Meanwhile, the single-layer Pb lattice behaves similarly to graphene monolayers in that it also generates a Dirac point at K, generating 3D massive Dirac fermions \cite{55}. Therefore, it is reasonable for the inherent quantum limit of the zeroth Landau level (LL) of the Dirac cone states in accordance with Abrikosov’s model of quantum magnetoresistance serves as a possible explanation for linear MR.
       
       In addition to the Dirac cone states, linear MR can also be explained by other mechanisms, particularly those involving broken spatial inversion symmetry. In materials where spatial inversion symmetry is broken, electron transport properties can differ from those in conventional materials, leading to a linear MR response. The distribution of defects or impurities can break specific symmetries, resulting in a non-uniform current distribution and thus inducing linear MR \cite{85}. Additionally, in topological insulators, surface states may contribute to resistivity in a manner that varies linearly with the magnetic field, especially when complex coupling between surface and bulk states is present. The existence of Dirac fermions, commonly found in topological insulators and certain high-temperature superconductors, can exhibit nontrivial magnetoresistance behavior, including linear MR, under strong magnetic fields due to their unique linear dispersion \cite{86}. Multiband effects can also lead to complex carrier scattering mechanisms and transport behavior in the presence of a magnetic field, which may result in linear MR \cite{87}. Furthermore, magnetic field-induced enhancement of carrier scattering, characterized by increased curvature of carrier trajectories under high magnetic fields, can enhance scattering effects. While this typically contributes to nonlinear MR behavior, under certain conditions, it may also result in linear MR \cite{88}.
       
        The WAL may originate from strong spin-orbit coupling (SOC) or spin momentum locking in the topological system. To get a deeper understanding of quantum WAL property, the study of magneto-conductance (MC) is highly necessary. The Hikami-Larkin-Nagaoka (HLN) equation is commonly employed to analyze the WAL effect \cite{51}. When considering the background with parabolic conductivity contribution, the modified HLN formula is given by:
      
       \begin{align}\label{eq1}
       	\Delta \sigma(B)=\alpha \frac{e^2}{2 \pi^2 \hbar}\left[\Psi\left(\frac{B_\phi}{B}+\frac{1}{2}\right)-\ln \left(\frac{B_\phi}{B}\right)\right] ,       \end{align}
where $\Psi$ represents the digamma function. $B_\phi$ = 1/(4e$L_{\phi}$$^2$) is the characteristic field with phase coherence length $L$$_\phi$=$\sqrt{D \tau}$, where $D$ is the diffusion constant and $\tau$ is the phase coherence time. The parameter $\alpha$ reflects the number of independent conducting channels causing interference. For WAL and weak localization (WL), the expected factors for $\alpha$ are -1/2 and 1, respectively. In systems exhibiting WAL effects due to spin-orbit coupling with parabolic dispersion and devoid of magnetic scattering (symplectic case), the value of $\alpha$ is -1/2. Similarly, in quantum spin-textured systems or topologically nontrivial conducting channels, each individual surface channel's $\alpha$ is also expected to be -1/2 owing to the existence of a nontrivial $\pi$ Berry phase. As a noncentrosymmetric superconductor, possessing a nonzero Z$_{2}$ topological invariant and topologically protected surface states, it can be predicted that the $\alpha$ value of PbTaSe$_{2}$ tends to -1/2 \cite{50}.
       
       Our data aligns well with the HLN fitting, as illustrated in Fig. 2(c). The extracted parameters $\alpha$ and $L_{\phi}$, are presented in Fig. 2(d). The modified HLN fitting at 5 K yields reasonable parameters $\alpha$ = - 0.45 which is slightly smaller than the theoretical value -0.5 expected for a single conduction channel. This indicates the presence of more than one conduction channel. For example, the presence of topologically trivial bands at the Fermi level could contribute to the conductivity in addition to the topological electrons. The value of $\alpha$ changes from $\sim$ -0.45 at 5 K to $\sim$ -0.05 at 30 K, consistent with that previously observed in other topological materials such as Cd$_{3}$As$_{2}$ \cite{73,53,75}. 
       
       The obtained $L$$_\phi$ decreases from 108 nm at 5 K to 32 nm at 30 K. Considering multiple scattering mechanisms, we try to analyze the temperature dependence of $L$$_\phi$ using the simple ­equation,
       \begin{align}\label{eq2}
      	\frac{1}{L_\phi^2}=\frac{1}{L_{\phi 0}^2}+ A_{\rm e e} T+ B_{\rm e p} T^p ,
       \end{align}
where $L_{\phi_{0}}$ represents the zero temperature dephasing length $A_{\rm e e} T$ and $B_{\rm e p}$ $T^{p}$ represent the contributions from electron-electron and electron-phonon scattering mechanisms, respectively \cite{80,81}. The value of $p$ for electron-phonon interaction in layered materials can vary between 2 to 3 depending on the effective dimensionality and disorder in the crystal \cite{77,78}. We obtained a good fit of $L$$_\phi$ with $p$ $\sim$ 2.5 as shown by the soild curves in Fig. 2(d). The extracted values of $A_{\rm e e}$ = 9.72 × 10$^{-6}$/nm$^{2}$K,  $B_{\rm e p}$ = 2.28 × 10$^{-7}$/nm$^{2}$K$^{2}$, and $L_{\phi0}$ $\approx$ 160.9 nm suggests the dominance of electron-electron scattering.
      
      As illustrated in Fig. 3, with the increase of disorder, the magnitude of the MR decreases. Furthermorer, the MR changes from the original V-type caused by WAL to an approximately linear MR (under high fields) in Figs. 3(a)-3(d). The shape change can be more easily observed in the first order derivative of MR shown in Figs. 3(f)-3(i). Initially, the slope increases with increasing magnetic field, but in the high-field region, it exhibits a trend towards stabilization, resembling a linear magnetoresistance. This behavior is akin to our previous research on FeSe, leading us to speculate that the linear MR is related to the Dirac cone state \cite{56}. Finally, as the irradiation dose increases up to 8$\times$10$^{16}$/cm$^{2}$, MR becomes a conventional quadratic type (Fig. 3(e)). Similarly, this phenomenon can be more easily observed in the first derivative (Fig. 3(j)). The plateau in the high-field region is suppressed, and the first derivative of the MR exhibits a completely linear change over the entire field range. The experiments confirm that there is a competitions between the WAL, linear MR ,and normal parabolic MR. 

	   Generally, the MR of irradiated PbTaSe$_{2}$ manifests a evolution from the WAL-dominant MR to a linear MR, and finally to a conventional parabolic MR. The MC change ratio $\Delta$MC is defined as $\Delta \mathrm{MC}=[\sigma(B)-\sigma(0)] / \sigma(0)$. MC can be expressed by $\sigma_T(B)=\sigma_{\mathrm{WAL}}+\sigma_{\mathrm{n}}$, where $\sigma_{\mathrm{WAL}}=a \sqrt{B}+\sigma_0$ is the surface conductivity from WAL corrections associated with intranode scattering, and $\sigma_{\mathrm{n}}=\left(\rho_0+C \cdot B+A \cdot B^2\right)^{-1}$ is from the contributions of linear and parabolic MR \cite{58}. The transition from a sharp decline to a parabolic shape in the fitting curve clearly indicates that with the increase in disorder, the WAL is suppressed.
	    
	    As shown in Fig. 4, we present the phase diagram of the coefficients $a$ with disorder. Below the critical temperature ($T$$_{c}$), the system exhibits superconductivity, with $T$$_{c}$ being almost independent of disorder. Above $T$$_{c}$, WAL is observed in the pristine single crystal. As temperature and disorder increase, WAL is gradually suppressed, leading to a regime where both WAL and linear MR coexist. In this regime, WAL is dominant at low temperatures, but as the temperature increases, linear MR becomes more apparent. The observed phenomenon can be attributed to the temperature-dependent behavior of electrons in the material. At relatively low temperatures, electrons primarily occupy the ground state. As the temperature rises, thermal excitations allow electrons to access higher energy states, leading to the suppression of quantum effects like WAL. Additionally, increased electron-phonon scattering at higher temperatures further diminishes WAL, leaving linear MR as the dominant effect. 

	   	\begin{figure}
	   	\centering
	   	\includegraphics[width=1.0\linewidth]{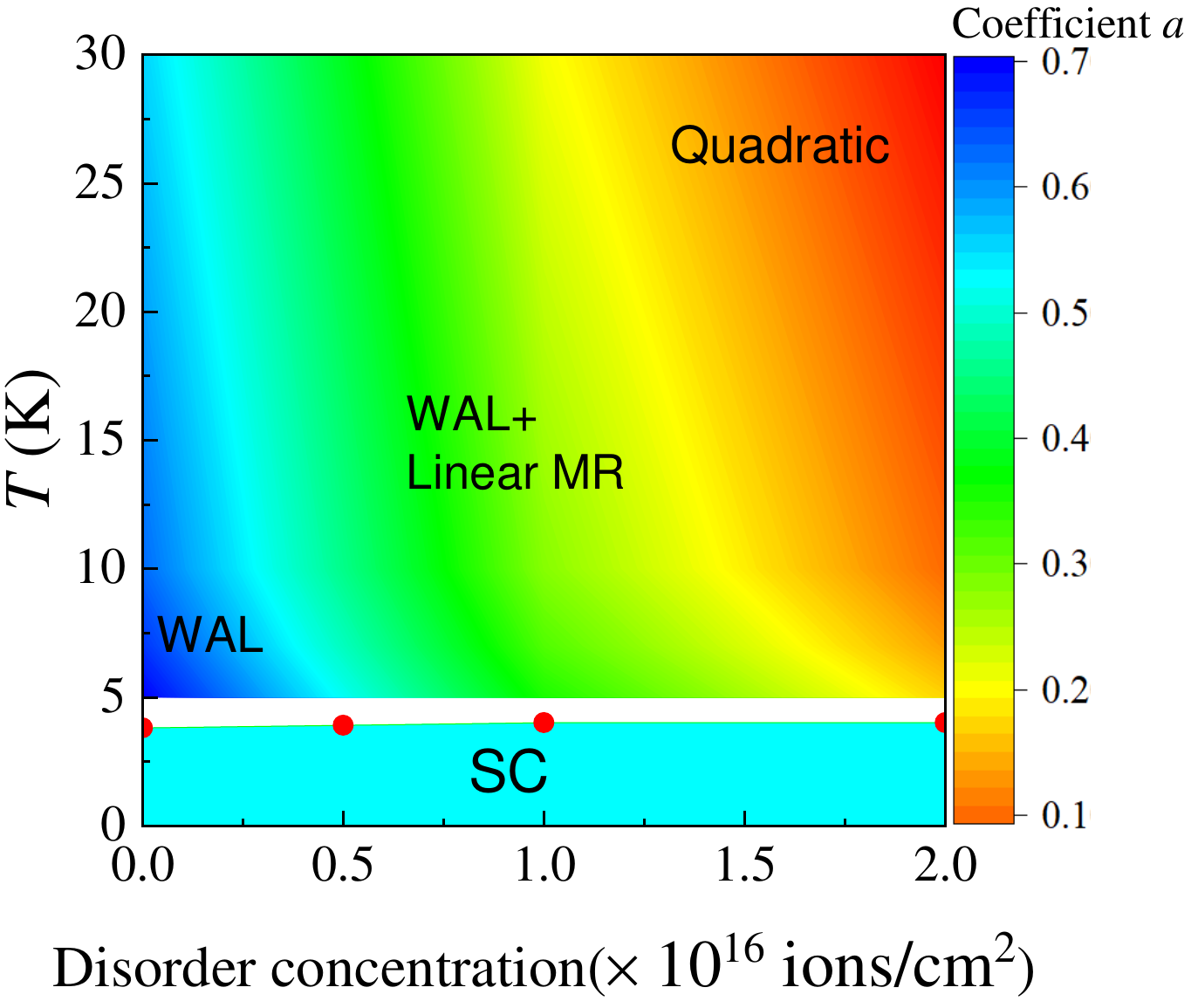}
	   	\caption{A phase diagram depicting the relationship between $\alpha$, disorder concentration, and temperature.}
	   	\label{fig:fig4}
	   \end{figure}
	   
       The introduction of disorders in topological materials, which enhances electron scattering within the lattice, typically leads to a decrease in MR. This inverse relationship between MR and disorder is a well-recognized phenomenon in topological semimetals: cleaner, less disordered systems tend to exhibit significantly higher MR, while increased impurity scattering in disordered systems disrupts electron trajectories, reducing carrier mobility and, consequently, diminishing MR. For examples, for the Dirac semimetal Cd$_{3}$As$_{2}$, higher-quality crystals show substantially larger MR, often reaching 10$^{5}$ percent at low temperatures and high magnetic fields \cite{82}. Similarly, in the Weyl semimetal TaAs, impurities and defects are known to reduce MR significantly, with clean crystals achieving MR values up to 10$^{6}$ \cite{83}. A comparable trend is observed in the Dirac semimetal ZrTe$_{5}$, where increased disorder leads to a marked reduction in MR \cite{84}. These illustrate a consistent pattern across various high-MR topological materials: reduced disorder corresponds to larger MR effects. The observed evolution in our study, from WAL-dominated MR to reduced MR with increasing impurity scattering, aligns with these findings, reinforcing the critical role of disorder control in the design and optimization of topological materials with enhanced MR properties. 
	
	   In conclusion, we have explored the magneto-transport properties of PbTaSe$_{2}$ under controlled disorder conditions, revealing a complex evolution of the MR. We observed a distinct crossover from WAL-dominated MR to a linear MR, and ultimately to a quadratic MR driven by classical Born scattering. These findings shed light on the intricate electronic behavior of PbTaSe$_{2}$ and underscore the interplay between various scattering mechanisms that shape its transport properties.

	   This work was partly supported by the National Key R\&D Program of China (Grant No. 2018YFA0704300), the National Natural Science Foundation of China (Grants No. 12374135, No. 12374136, No. 12204265, No. 12204487), the Ministry of Science and Technology in Taiwan under Projects No. NSTC-111-2124-M-001-007, No. NSTC-1102112-M-001-065-MY3, and No.NSTC111-2124-M-A49-003 and by Academia Sinica for the budget of AS-iMATE-11312.
	    
	    L.S. Y.S. and Q.H.contributed equally to this paper.

	\acknowledgements
	
\bibliographystyle{unsrt}%
\bibliography{ref}

\end{document}